# Electrostatic multipole contributions to the binding energy of electrons

A. D. Alhaidari[(a)] and H. Bahlouli[(b)]

[(a)] *Saudi Center for Theoretical Physics, P.O. Box 32741, Jeddah 21438, Saudi Arabia*

[(b)] *Physics Department, King Fahd University of Petroleum & Minerals, Dhahran 31261, Saudi Arabia*

**Abstract:** The interaction of an electron with a local static charge distribution (e.g., an atom or molecule) is dominated at large distances by the radial $1/r$ Coulomb potential. The second order effect comes from the non-central electric dipole contribution $\cos\theta/r^2$. Moreover, the third order effect is due to the electric quadrupole potential, $(3\cos^2\theta - 1)/2r^3$. We use the tridiagonal representation approach to give a reasonably accurate account for the combined effects of all these contributions to the binding energy of the electron but with an effective quadrupole interaction. As an application, we obtain the bound states of a valence electron in an atom with both electric dipole and quadrupole moments.

## 1. Introduction

In the atomic units, $\hbar = M = e = 1$, the time-independent three-dimensional Schrödinger equation for an electron (mass $M$ and charge $-e$) in the field of a static charge distribution described by an electrostatic potential function $V(\vec{r})$ reads as follows

$$\left[-\tfrac{1}{2}\vec{\nabla}^2 + V(\vec{r}) - E\right]\psi(\vec{r}) = 0, \tag{1}$$

where $\vec{\nabla}^2$ is the three-dimensional Laplacian, $E$ is the electron energy and $\psi(\vec{r})$ is the associated wavefunction. In spherical coordinates, the potential function $V(\vec{r})$ associated with the local charge distribution at distances much larger than the size of the distribution can well be represented by a multipole expansion, which when written up to and including the linear electric quadrupole, reads as follows (see, for example, Ref. [1])

$$V(\vec{r}) = -\frac{Q}{r} - d\frac{\cos\theta}{r^2} + q\frac{\tfrac{1}{2}(3\cos^2\theta - 1)}{r^3}, \tag{2}$$

where $Q$ is the net effective positive nuclear charge felt by the valence electron, $d$ is the electric dipole moment along the positive $z$-axis, and $q$ is the linear electric quadrupole moment of the charge distribution. We took the Bohr radius, $a_0 = 4\pi\varepsilon_0\hbar^2/Me^2 = 4\pi\varepsilon_0$, as the unit of length. Now, the quadrupole term in the potential (2) destroys separability of the wave equation (1) making its solution a highly non-trivial task. Therefore, we consider an effective electric quadrupole interaction where the angular factor $\tfrac{1}{2}(3\cos^2\theta - 1)$ is replaced by a dimensionless angular parameter $\eta$ such that $-\frac{1}{2} \leq \eta \leq +1$ since $0 \leq \cos^2\theta \leq 1$. This could be considered as being some average over the angular dependence and results in an effective quadrupole potential $p/r^3$, where the effective electric quadrupole moment is $p = \eta q$. Consequently, the radial part of the wave equation (1) becomes

$$\left[ -\frac{1}{2}\frac{d^2}{dr^2} + \frac{\gamma(\gamma+1)}{2r^2} - \frac{Q}{r} + \frac{p}{r^3} - E \right]\psi(r) = 0, \tag{3}$$

where $\gamma$ is a quantum number that depends on the electric dipole moment $d$ and the azimuthal angular momentum quantum number $m = 0, \pm 1, \pm 2, \ldots$ [2]. If $d = 0$ then $\gamma$ becomes the orbital angular momentum quantum number $\ell = 0, 1, 2, \ldots$. On the other hand, for a non-zero dipole moment, Eq. (3.6) and Eq. (3.8) in Ref. [2] give $\left(\gamma + \frac{1}{2}\right)^2$ as one of the eigenvalues of an infinite symmetric tridiagonal matrix whose elements are given by

$$T_{i,j} = \left(i + m + \tfrac{1}{2}\right)^2 \delta_{i,j} - d\sqrt{\tfrac{i(i+2m)}{(i+m)^2 - 1/4}}\, \delta_{i,j+1} - d\sqrt{\tfrac{(i+1)(i+2m+1)}{(i+m+1)^2 - 1/4}}\, \delta_{i,j-1}. \tag{4}$$

For pure dipolar interaction ($p = 0$), the condition $\left(\gamma + \frac{1}{2}\right)^2 > 0$ eliminates quantum anomalies associated with the orbital inverse square potential [3]. However, for $p > 0$ this condition is not required for a physical solution.

In section 2, we start by formulating the problem using the "tridiagonal representation approach (TRA)". In section 3, we obtain the TRA solution for non-zero positive net charge $Q$. In section 4, we present numerical results where we show the effect of the pure quadrupole ($d = 0$) contribution to the electron binding energy after filtering out the Coulomb (electric monopole) contribution. In section 5, a realistic model is presented where we obtain the bound states of a valence electron in an atom with electric dipole and quadrupole moments. Then, we conclude in section 6 with some relevant remarks.

## 2. TRA formulation of the problem

The combined $r^{-1}$, $r^{-2}$, and $r^{-3}$ power-law potential in Eq. (3) for arbitrary physical parameters $\{\gamma, Q, p\}$ has no known exact solution in the published literature. Nonetheless, in this work we give a highly accurate *analytic* approximation in a properly chosen finite basis set. To avoid quantum anomalies due to the most singular term in the potential, the effective electric quadrupole interaction term $pr^{-3}$, we require that the angular parameter $\eta$ be chosen such that $p = \eta q$ is positive. We employ the tools of the TRA in the formulation and solution of this problem. For details on the TRA and how it is used in solving the wave equation, one may consult [4,5] and references cited therein. We start by expanding the wavefunction in a pointwise convergent series as $\psi(r) = \sum_n f_n \phi_n(x)$, where $\{\phi_n(x)\}$ is a set of square integrable functions that produce a tridiagonal matrix representation for the wave operator (3). We choose the dimensionless variable $x$ as $x = 1/\lambda r$, where $\lambda$ is an arbitrary real positive scale parameter with inverse length dimension. In terms of this variable, the wave equation (3) becomes

$$\mathcal{D}\psi(r) = -\frac{\lambda^2 x^2}{2}\left[x^2\frac{d^2}{dx^2} + 2x\frac{d}{dx} - \gamma(\gamma+1) + \frac{2Q/\lambda}{x} - (2\lambda p)x + \frac{\varepsilon}{x^2}\right]\psi(r) = 0. \tag{5}$$

where $\varepsilon = 2E/\lambda^2$. A proper choice of square-integrable basis that could support a tridiagonal matrix representation for the wave operator $\mathcal{D}$ has the following elements

$$\phi_n(x) = x^\alpha e^{-1/2x} Y_n^\mu(x), \tag{6}$$

where $Y_n^\mu(x)$ is the Bessel polynomial on the positive real line whose relevant properties are shown in the Appendix and $\mu < -\frac{1}{2}$. The degree of the polynomial is limited by the negative parameter $\mu$ as $n = 0, 1, 2, ..., N$ and $N$ being the largest integer less than $-\mu - \frac{1}{2}$. Therefore, the basis set $\{\phi_n(x)\}_{n=0}^N$ is finite. Consequently, this basis can produce a faithful physical representation for the bound states of a system whose discrete spectrum is also finite and with a maximum size of $N+1$. Nonetheless, it can also support a quasi-exact solution (finite portion of the spectrum) for a system with an infinite discrete spectrum and with an accuracy that increases with the size of the basis. We will shortly discover that for a fixed positive net charge $Q$, the size of the basis for the system under consideration increases with the bound state energy level. This is consistent with the fact that higher excited states have larger number of nodes (oscillation) requiring a larger degree of the polynomial $Y_n^\mu(x)$ for accurate representation, which means larger basis size.

Using the differential equation of the Bessel polynomial (A4) in the Appendix and choosing $\alpha = \mu$, we can evaluate the action of the wave operator on the basis elements giving

$$\mathcal{D}\phi_n(x) = -\frac{\lambda^2}{2}x^{\mu+2}e^{-1/2x}\left[\left(n+\mu+\tfrac{1}{2}\right)^2 - \left(\gamma+\tfrac{1}{2}\right)^2 + \frac{\mu+2Q/\lambda}{x} - (2\lambda p)x + \frac{\varepsilon+1/4}{x^2}\right]Y_n^\mu(x). \tag{7}$$

A tridiagonal representation of the wave operator $\mathcal{D}$ is obtained if and only if the right-hand side of Eq. (7) becomes a sum of terms proportional to $\phi_n(x)$ and $\phi_{n\pm 1}(x)$ with constant factors (modulo an overall multiplicative function that does not depend on $n$). That is, [4,5]

$$\mathcal{D}\phi_n(x) = \omega(x)\left[a_n \phi_n(x) + b_{n-1}\phi_{n-1}(x) + c_n \phi_{n+1}(x)\right], \tag{8}$$

where $\{a_n, b_n, c_n\}$ are $x$-independent parameters and $\omega(x)$ is a node-less entire function. Consequently, the three-term recursion relation (A2) dictates that the expression inside the square brackets in (7) must be linear in $x$. Therefore, the terms proportional to $x^{-2}$ and $x^{-1}$ must vanish and thus we must choose the basis parameters $\lambda$ and $\mu$ as follows

$$\lambda = 2\sqrt{-2E}, \tag{9a}$$

$$\mu = -Q/\sqrt{-2E}. \tag{9b}$$

Hence, our TRA solution will be restricted to negative energies (i.e., bound states). Moreover, since $\mu$ is negative then the net charge of the electrostatic distribution, $Q$, seen by the electron must be positive. Additionally, since $N$ is the largest integer less than $-\mu - \frac{1}{2}$, then relation (9b) implies that the size of the basis increases as $|E|$ becomes smaller (i.e., as the energy level gets higher) as noted above.

In the following section, we present the TRA solution of the problem which is written in terms of a new orthogonal polynomial defined in [6] by its three-term recursion relation and initial values.

## 3. TRA solution of the problem

With the basis parameters $\mu$ and $\lambda$ given by (9), the action of the wave operator (7) on the basis elements reduces to the following

$$\mathcal{D}\phi_n(x) = 4E\, x^{\mu+2} e^{-1/2x} \left[ \left(n+\mu+\tfrac{1}{2}\right)^2 - \left(\gamma+\tfrac{1}{2}\right)^2 - (2\lambda p)x \right] Y_n^\mu(x). \tag{10}$$

Substituting this action in the wave equation $\mathcal{D}\psi(r) = \sum_n f_n \mathcal{D}\phi_n(x) = 0$ and using the recursion relation of the Bessel polynomial (A2), we obtain the following three-term recursion relation for the expansion coefficients

$$\left(\gamma+\tfrac{1}{2}\right)^2 F_n = \left[ \left(n+\mu+\tfrac{1}{2}\right)^2 + \frac{\mu\lambda p}{(n+\mu)(n+\mu+1)} \right] F_n$$
$$+ \lambda p \left[ \frac{n+1}{(n+\mu+1)(2n+2\mu+3)} F_{n+1} - \frac{n+2\mu}{(n+\mu)(2n+2\mu-1)} F_{n-1} \right] \tag{11}$$

where we have written $f_n = f_0 F_n$ making $F_0 = 1$. If we define $P_n = \dfrac{(-1)^n n!(2\mu+1)}{(2n+2\mu+1)(2\mu+1)_n} F_n$, where $(a)_n = a(a+1)(a+2)\ldots(a+n-1) = \dfrac{\Gamma(n+a)}{\Gamma(a)}$ is the Pochhammer symbol (a.k.a. shifted factorial), then this recursion becomes

$$\left(\gamma+\tfrac{1}{2}\right)^2 P_n = \left[ \left(n+\mu+\tfrac{1}{2}\right)^2 + \frac{\mu\lambda p}{(n+\mu)(n+\mu+1)} \right] P_n$$
$$+ \frac{\lambda p}{2} \left[ \frac{n}{(n+\mu)\left(n+\mu+\tfrac{1}{2}\right)} P_{n-1} - \frac{n+2\mu+1}{(n+\mu+1)\left(n+\mu+\tfrac{1}{2}\right)} P_{n+1} \right] \tag{12}$$

Comparing this recursion relation to that of the polynomial $B_n^\mu(z;\sigma)$ defined in [6] by its three-term recursion relation and initial values, which is shown here in the Appendix as (A10), we conclude that $P_n = B_n^\mu(z;\sigma)$ with

$$\sigma = -1\big/p\sqrt{-2E}, \qquad z = -\left(\gamma+\tfrac{1}{2}\right)^2\big/p\sqrt{-2E}. \tag{13}$$

Moreover, the k-th bound state wavefunction for the electron reads

$$\psi_k(r) \approx f_0(E_k)(\lambda r)^{-\mu} e^{-\lambda r/2} \sum_{n=0}^{N} G_n B_n^\mu(z;\sigma) Y_n^\mu(1/\lambda r), \tag{14}$$

where $G_n = (2n+2\mu+1)(2\mu+1)_n\big/(-1)^n n!(2\mu+1)$. For a given set of physical parameters and bound state energy $E_k$, the basis parameters $\mu$ and $\lambda$ are given by (9) whereas $z$ and $\sigma$ are given by (13). Therefore, to have a full representation of the wavefunction (14), we only need an evaluation of the corresponding energy $E_k$. Now, all properties of the system including the energy spectrum of the bound states are obtained from the properties of the polynomial $B_n^\mu(z;\sigma)$ (weight function, generating function, asymptotics, zeros, etc.). Unfortunately, the analytic properties of $B_n^\mu(z;\sigma)$ are not yet known. It remains an open problem in orthogonal polynomials along with other similar problems. For an exposé of these open problems, one may consult [7,8] and references therein. Therefore, we are forced to resort to numerical means to calculate the energy spectrum of the bound states.

In the following section, we use a robust numerical scheme based on Gauss quadrature and continued fractions to calculate the bound states energy eigenvalues $\{E_k\}$ for a given set of physical parameters $\{Q,d,q\}$, azimuthal quantum number $m$ and angular parameter $\eta$. In the next section we will start by considering the quadrupole effect alone, that is, we limit our calculation to the special case where $d=0$. Hence, we obtain the combined effect of the electric monopole (Coulomb potential) and quadrupole on the binding energy then we filter out the Coulomb part to get an evaluation of the quadrupole contribution. However, in section 5, we consider the combined contributions of the electric dipole and quadrupole moments of a given atom to the binding energy of the valence electrons.

## 4. Binding energy calculation

Since the analytic properties of the TRA polynomial $B_n^\mu(z;\sigma)$ that appears in the wavefunction expansion (14) are not known, we might be tempted to use the symmetric version of the recursion relation (12) to obtain the energy spectrum. To do that, we rewrite (12) as an eigenvalue equation $t|P\rangle = \mathcal{T}|P\rangle$ where $t = \left(\gamma + \tfrac{1}{2}\right)^2$ and $\mathcal{T}$ is a symmetric tridiagonal matrix with elements $\mathcal{T}_{n,m} = a_n \delta_{n,m} + b_{n-1} \delta_{n,m+1} + b_n \delta_{n,m-1}$, where

$$a_n = \left(n + \mu + \tfrac{1}{2}\right)^2 + \frac{\mu \lambda p}{(n+\mu)(n+\mu+1)}, \tag{15a}$$

$$b_n = \frac{\lambda p}{n+\mu+1} \sqrt{\frac{-(n+1)(n+2\mu+1)}{(2n+2\mu+1)(2n+2\mu+3)}}. \tag{15b}$$

The problem in finding the energy spectrum this way is that the matrix elements of $\mathcal{T}$ do depend on the energy itself through $\mu$ and $\lambda$. However, out of all energies that enter in the construction of $\mathcal{T}$, the allowed energies are only those that produce the eigenvalue $t = \left(\gamma + \tfrac{1}{2}\right)^2$ for a given $\gamma$. The technique that utilizes this idea to produce the energy spectrum is called the "potential parameter spectrum" (PPS) [9]. Unfortunately, it has been demonstrated elsewhere that the PPS does not produce accurate enough results in a finite basis settings like the current basis (6). Therefore, we present an alternative robust numerical scheme for evaluating the energy spectrum as follows.

We start by selecting a proper complete square integrable basis then evaluate the matrix elements of the Hamiltonian operator in this basis. Thereafter, we can obtain the energy spectrum by diagonalizing this matrix and keeping only the negative eigenvalues. For $d=0$ and $\gamma = \ell$ we choose the "Laguerre basis" whose elements are defined by

$$\chi_n(y) = A_n y^{\ell+1} e^{-y/2} L_n^{2\ell+1}(y), \tag{16}$$

where $y = \rho r$ and $L_n^{2\ell+1}(y)$ is the Laguerre polynomial with $\rho$ being a positive scale parameter of inverse length dimension. The normalization constant is chosen as $A_n = \sqrt{n!/(n+2\ell+1)!}$. The action of the Hamiltonian operator, $H = -\frac{1}{2}\frac{d^2}{dr^2} + V(r)$, on the basis (16) gives

$$H\chi_n(y) = \frac{\rho^2}{2} A_n y^{\ell} e^{-y/2}\left[(n+\ell+1) - 2\frac{Q}{\rho} - \frac{y}{4}\right] L_n^{2\ell+1}(y) + p\frac{\rho^3}{y^3}\chi_n(y), \tag{17}$$

where we have utilized the differential equation of the Laguerre polynomial. Now, we use the recursion relation of the Laguerre polynomials, $yL_n^\nu(y) = (2n+\nu+1)L_n^\nu(y) - (n+\nu)L_{n-1}^\nu(y) - (n+1)L_{n+1}^\nu(y)$, for the last term inside the square brackets in (17). The result is as follows

$$\begin{aligned}H\chi_n(y) &= p\frac{\rho^3}{y^3}\chi_n(y) + \frac{\rho^2}{4y}\left[(n+\ell+1) - 4\frac{Q}{\rho}\right]\chi_n(y) \\ &+ \frac{\rho^2}{8y}\left[\sqrt{n(n+2\ell+1)}\,\chi_{n-1}(y) + \sqrt{(n+1)(n+2\ell+2)}\,\chi_{n+1}(y)\right]\end{aligned} \tag{18}$$

Using the orthogonality relation of the Laguerre polynomials, $\int_0^\infty y^\nu e^{-y} L_n^\nu(y) L_m^\nu(y) dy = \frac{\Gamma(n+\nu+1)}{\Gamma(n+1)}\delta_{n,m}$, we obtain the following elements of the Hamiltonian matrix

$$\begin{aligned}\langle \chi_n | H | \chi_m \rangle &= \rho^3 p \langle n | y^{-2} | m \rangle + \frac{\rho^2}{4}\left[(n+\ell+1) - 4\frac{Q}{\rho}\right]\delta_{n,m} \\ &+ \frac{\rho^2}{8}\left[\sqrt{n(n+2\ell+1)}\,\delta_{n,m+1} + \sqrt{(n+1)(n+2\ell+2)}\,\delta_{n,m-1}\right]\end{aligned} \tag{19}$$

where we have defined $\langle n | y^{-2} | m \rangle := A_n A_m \int_0^\infty y^{2\ell-1} e^{-y} L_n^{2\ell+1}(y) L_m^{2\ell+1}(y) dy$. This integral could be evaluated numerically using Gauss quadrature integral approximation associated with the Laguerre polynomial (see, for example, Appendix B in [10] or Appendix A in [11]). Rigorously, integrability dictates that the exponent $2\ell-1$ of $y$ inside the integral be greater than $-1$. Thus, we expect integration difficulty only for $\ell = 0$. However, numerically we can improve accuracy of the integration result by increasing the size of the basis (16) and/or choosing a proper value for the arbitrary scale parameter $\rho$. The energy spectrum $\{E_k\}$ is identified as the set of negative eigenvalues of the following generalized eigenvalue matrix wave equation

$$\mathcal{H}|\psi_k\rangle = E_k \Omega |\psi_k\rangle, \tag{20}$$

where $\mathcal{H}$ is the Hamiltonian matrix whose elements are given by (19) and $\Omega$ is the overlap matrix of the basis (16) whose elements are defined by

$$\begin{aligned}\Omega_{n,m} &= \langle \chi_n | \chi_m \rangle = A_n A_m \int_0^\infty y^{2\ell+2} e^{-y} L_n^{2\ell+1}(y) L_m^{2\ell+1}(y) dy = \langle n | y | m \rangle \\ &= 2(n+\ell+1)\delta_{n,m} - \sqrt{n(n+2\ell+1)}\,\delta_{n,m+1} - \sqrt{(n+1)(n+2\ell+2)}\,\delta_{n,m-1}\end{aligned} \tag{21}$$

In Table 1, we give the results of the above procedure in the form of the energy deviations $E_k - E_k^C$, where $E_k^C = -Q^2/2(k+\ell+1)^2$ is the Coulomb interaction energy. It is clear from the Table that this deviation is significant for low angular momenta and low excitations indices.

It should be noted that the overlap matrix $\Omega$ does not depend on the non-physical computational parameter $\rho$ but the Hamiltonian matrix $\mathcal{H}$ does. However, this dependence is a numerical artifact in the sense that the computed energy eigenvalues will be independent of $\rho$ if the size of the truncated matrices used in the calculation becomes infinite. For finite sizes, however, we search for a range of values of $\rho$ where the eigenvalues do not change (within the desired accuracy) as we vary $\rho$ in this range. This range is called the "plateau of stability" for $\rho$ whose width increases with the size of the matrices. Theoretically, the width of the plateau goes to infinity as the size of the matrices go to infinity. Our choice of $\rho$ for the results shown in the Table is taken from the middle of the stability plateau, $\rho = 2$..

With the energy spectrum $\{E_k\}$ obtained, all ingredients needed to compute the bound states wavefunction (14) are determined. Figure 1 is a plot of the lowest energy states corresponding to the physical parameters in Table 1 for $\ell = 0$ and $\ell = 1$. Figure 2 is a set of plots of the wavefunctions in the finite basis (6) as given analytically by the series (14) shown in blue superimposed by those calculated numerically in the complete basis (16) shown in red for $\ell = 1$. The figure confirms our observation made in section 2 that the size of the basis (6) increases with the bound state energy level resulting in a more accurate representation of the wavefunction. The size of the basis (6) corresponding to the states shown in Figure 2 from $\psi_0(r)$ to $\psi_{10}(r)$ are: 3, 4, 5, 7, 10, and 13. That is, $N = k + 2$, where $k$ is the energy level index.

## 5. Binding of valence electron

As an interesting application of the above findings, we compute the bound states (energy spectrum and wavefunction) for an electron in the valence band of an atom/molecule with both electric dipole and quadrupole moments. In large atoms with a single electron in the valence band, the valence electron is exceptionally far from the nucleus and the rest of the electrons in the suborbitals. This makes the multipole expansion (2) for the total atomic charge minus that of the valence electron (i.e., $Q = +1$) well justified. Thus, our result is expected to be more accurate for such valence electrons. On the other hand, if the valence band contains two electrons, then better results are obtained if one of the valence electrons is stripped away from the atom leaving a positively charged atom (cation[+]) where we set $Q = +2$. And so on: cation[++] with three electrons in the valence band, two of which were stripped away making $Q = +3$, etc.

For a given atom/molecule with moments $\{d, q\}$, azimuthal quantum number $m$ and angular parameter $\eta$, we use the numerical scheme in section 4 to obtain the bound state energies $\{E_k\}$ with $Q = +1$ as shown in Table 2 [12]. Figure 3 shows the un-normalized wavefunctions in the finite basis (6) in blue as given analytically by the series (14) super-imposed by those calculated numerically in the complete basis (16) in red. We used the parameter values in Table 2 with states corresponding to $m = 1$ and $\gamma = 1.973702871$.

# 6. Conclusion

In the present work we computed the binding energy of an electron that is loosely bound to an atom or molecule represented by a local static electric charge distribution. The electron potential energy associated with this interaction is approximated by an electrostatic multipole expansion which includes, monopole, dipole and quadrupole contributions simultaneously. However, the angular distribution of the quadrupole was replaced by an effective angular parameter making the quadrupole contribution to the electrostatic potential radial with inverse cube power.

To find the eigenvalues of the Hamiltonian, we used the TRA in a finite basis to obtain an approximation for the bound states of a valence electron located far away from a static charge distribution with a net effective positive charge. Contrary to the traditional approximation schemes where the lowest bound states are most accurate, we find that our approximation improves for higher excited states. As an application, we considered the bound states of the valence electron in an atom/molecule where the valance band contains a single electron. If the band contains more than one electron then our findings apply if the excess electrons in the band are stripped off forming a cation. However, it is unfortunate that we could not make a realistic comparison with experimental results on quadrupole bound states because of the scarcity of such data on loosely bound electrons.

In another complementary study, we used the TRA to obtain solutions of an electron scattering off a neutral charge distribution (e.g., atom or molecule) with electric dipole and quadrupole moments [13]. In that study, the basis size was infinite and written in terms of the Bessel function with discrete index, $J_{n+\nu}(x)$. This is in contrast to the finite basis in the present study, which is written in terms the of the Bessel polynomials $Y_n^\mu(x)$.

Finally, we need to stress that while the multipole expansion we used for our electrostatic model describes rigorously the interaction of the valence electron with the molecular charge distribution at large distances and for large atoms/molecules, it breaks down once this electron enters the charge cloud of the corresponding atom or molecule, a phenomenon called charge penetration effect [16].

# Acknowledgements

We are grateful to I. A. Assi for conducting an independent verification of our results in the Tables using the "Lagrange mesh" method and the "discrete variable representation" method.

# Appendix: Bessel polynomials on the real line

For ease of reference to the reader, we reproduce in this Appendix (with kind permission of The European Physical Journal) all relevant properties of the Bessel polynomial found in our work in [14]. These polynomials are defined in terms of the hypergeometric or confluent hypergeometric functions as follows (see section 9.13 of the book by Koekoek *et. al* [15] but with the replacement $x \mapsto 2x$ and $a \mapsto 2\mu$)

$$Y_n^\mu(x) = {}_2F_0\left(\begin{matrix}-n, n+2\mu+1\\ \text{—}\end{matrix}\middle| -x\right) = (n+2\mu+1)_n\, x^n\, {}_1F_1\left(\begin{matrix}-n\\-2(n+\mu)\end{matrix}\middle| 1/x\right), \tag{A1}$$

where $x \geq 0$, $n = 0,1,2,..,N$ and $N$ is the largest non-negative integer less than $-\mu - \frac{1}{2}$. The Pochhammer symbol $(a)_n$ (a.k.a. shifted factorial) is defined as $(a)_n = a(a+1)(a+2)...(a+n-1) = \frac{\Gamma(n+a)}{\Gamma(a)}$. The Bessel polynomial could also be written in terms of the Laguerre polynomial as: $Y_n^\mu(x) = n!(-x)^n L_n^{-(2n+2\mu+1)}(1/x)$. The three-term recursion relation reads as follows:

$$2xY_n^\mu(x) = \frac{-\mu}{(n+\mu)(n+\mu+1)} Y_n^\mu(x)$$
$$- \frac{n}{(n+\mu)(2n+2\mu+1)} Y_{n-1}^\mu(x) + \frac{n+2\mu+1}{(n+\mu+1)(2n+2\mu+1)} Y_{n+1}^\mu(x) \tag{A2}$$

Note that the constraints on $\mu$ and on the maximum polynomial degree make this recursion definite (i.e., the values of the two recursion coefficients multiplying $Y_{n\pm1}^\mu(x)$ have the same sign). Otherwise, these polynomials could not be defined on the real line but on the unit circle in the complex plane. The orthogonality relation reads as follows

$$\int_0^\infty x^{2\mu} e^{-1/x} Y_n^\mu(x) Y_m^\mu(x) dx = -\frac{n!\Gamma(-n-2\mu)}{2n+2\mu+1} \delta_{nm}. \tag{A3}$$

The differential equation is

$$\left\{ x^2 \frac{d^2}{dx^2} + [1 + 2x(\mu+1)] \frac{d}{dx} - n(n+2\mu+1) \right\} Y_n^\mu(x) = 0. \tag{A4}$$

The forward and backward shift differential relations read as follows

$$\frac{d}{dx} Y_n^\mu(x) = n(n+2\mu+1) Y_{n-1}^{\mu+1}(x). \tag{A5}$$

$$x^2 \frac{d}{dx} Y_n^\mu(x) = -(2\mu x + 1) Y_n^\mu(x) + Y_{n+1}^{\mu-1}(x). \tag{A6}$$

We can write $Y_{n+1}^{\mu-1}(x)$ in terms of $Y_n^\mu(x)$ and $Y_{n\pm1}^\mu(x)$ as follows

$$2Y_{n+1}^{\mu-1}(x) = \frac{(n+1)(n+2\mu)}{(n+\mu)(n+\mu+1)} Y_n^\mu(x)$$
$$+ \frac{n(n+1)}{(n+\mu)(2n+2\mu+1)} Y_{n-1}^\mu(x) + \frac{(n+2\mu)(n+2\mu+1)}{(n+\mu+1)(2n+2\mu+1)} Y_{n+1}^\mu(x) \tag{A7}$$

Using this identity and the recursion relation (A2), we can rewrite the backward shift differential relation as follows

$$2x^2 \frac{d}{dx} Y_n^\mu(x) = n(n+2\mu+1) \times$$
$$\left[ -\frac{Y_n^\mu(x)}{(n+\mu)(n+\mu+1)} + \frac{Y_{n-1}^\mu(x)}{(n+\mu)(2n+2\mu+1)} + \frac{Y_{n+1}^\mu(x)}{(n+\mu+1)(2n+2\mu+1)} \right] \tag{A8}$$

The generating function is

$$\sum_{n=0}^{\infty} Y_n^{\mu}(x)\frac{t^n}{n!} = \frac{2^{2\mu}}{\sqrt{1-4xt}}\left(1+\sqrt{1-4xt}\right)^{-2\mu}\exp\left[2t/(1+\sqrt{1-4xt})\right]. \tag{A9}$$

The polynomial $B_n^{\mu}(z;\sigma)$ that appears in the wavefunction expansion (14) is defined in [6] by its three-term recursion relation Eq. (15) therein, which reads

$$z B_n^{\mu}(z;\sigma) = \left[\frac{-2\mu}{(n+\mu)(n+\mu+1)} + \sigma\left(n+\mu+\tfrac{1}{2}\right)^2\right]B_n^{\mu}(z;\sigma)$$
$$-\frac{n}{(n+\mu)\left(n+\mu+\tfrac{1}{2}\right)}B_{n-1}^{\mu}(z;\sigma) + \frac{n+2\mu+1}{(n+\mu+1)\left(n+\mu+\tfrac{1}{2}\right)}B_{n+1}^{\mu}(z;\sigma) \tag{A10}$$

where $B_0^{\mu}(z;\sigma)=1$ and $B_{-1}^{\mu}(z;\sigma):=0$.

# References

[1] D. J. Griffiths, *Introduction to electrodynamics*, 4th ed. (Pearson Education, 2013) Section 3.4

[2] A. D. Alhaidari, *Analytic solution of the wave equation for an electron in the field of a molecule with an electric dipole moment*, Ann. Phys. **323** (2008) 1709

[3] A. M. Essin and D. J. Griffiths, *Quantum mechanics of the 1/x^2 potential*, Am. J. Phys. **74** (2006) 109

[4] A. D. Alhaidari, *Solution of the nonrelativistic wave equation using the tridiagonal representation approach*, J. Math. Phys. **58** (2017) 072104

[5] A. D. Alhaidari and H. Bahlouli, *Tridiagonal Representation Approach in Quantum Mechanics*, Phys. Scripta **94** (2019) 125206

[6] A.D. Alhaidari, *Exponentially confining potential well*, Theor. Math. Phys. **206** (2021) 84

[7] A. D. Alhaidari, *Open problem in orthogonal polynomials*, Rep. Math. Phys. **84** (2019) 393

[8] W. Van Assche, *Solution of an open problem about two families of orthogonal polynomials*, SIGMA **15** (2019) 005

[9] A. D. Alhaidari and H. Bahlouli, *Bound states and the potential parameter spectrum*, J. Math. Phys. **61** (2020) 062103

[10] A. D. Alhaidari, H. Bahlouli, C. P. Aparicio, and S. M. Al-Marzoug, *J-matrix method of scattering for inverse-square singular potentials with supercritical coupling I. No regularization*, Ann. Phys. **445** (2022) 169020

[11] A. D. Alhaidari, *Reconstructing the potential function in a formulation of quantum mechanics based on orthogonal polynomials*, Commun. Theor. Phys. **68** (2017) 711

[12] In this calculation, the numerical scheme of section 4 should be modified by the replacement $\ell \mapsto \gamma$ where $\gamma$ is obtained from the eigenvalues $\left(\gamma+\tfrac{1}{2}\right)^2$ of the tridiagonal

symmetric matrix (4). Note however, that if $\left(\gamma+\tfrac{1}{2}\right)^2 < 0$ and $p > 0$ then $\gamma = -\tfrac{1}{2} + i\omega$ where $\omega$ is real and, in this case, the numerical scheme does not work. In Table 2, two such cases occur that are not shown: one for $m = 0$ and another for $m = 1$.

[13] A. D. Alhaidari and M. E. H. Ismail, *Solutions of the scattering problem in a complete set of Bessel functions with a discrete index*, arXiv:2209.03738 [quant-ph]

[14] A. D. Alhaidari, I. A. Assi, and A. Mebirouk, *Bound states of a quartic and sextic inverse power-law potential for all angular momenta,* Eur. Phys. J. Plus **136** (2021) 443: With kind permission of The European Physical Journal.

[15] R. Koekoek, P. A. Lesky and R. F. Swarttouw, *Hypergeometric Orthogonal Polynomials and Their q-Analogues* (Springer, Heidelberg, 2010)

[16] J. A. Rackers, Q. Wang, C. Liu, J.-P. Piquemal, P. Ren and J. W. Ponder, *An optimized charge penetration model for use with the AMOEBA force field*, Phys. Chem. Chem. Phys. **19** (2017) 276

## Table Caption:

**Table 1**: The lowest binding energies (in atomic units) of an electron to the local charge distribution after subtracting the Coulomb contribution. We took $Q = 2$, $p = \eta q = 5$ and several values of the angular momentum. We used the Hamiltonian matrix diagonalization (HMD) method in the Laguerre basis as explained in section 4. Matrices are 150×150 and the scale parameter $\rho$ is chosen from the middle of the plateau of stability as $\rho = 2$.

**Table 2**: The lowest binding energies (in atomic units) of a valence electron in a neutral atom ($Q = 1$) with electric dipole and quadrupole moments $d = 5$ and $p = \eta q = 3$. The values of $\gamma$ are obtained from the eigenvalues of the tridiagonal symmetric matrix (4). HMD numerical parameters are $N = 150$ and $\rho = 2$.

## Figure Caption:

**Fig. 1**: The un-normalized bound states corresponding to the lowest part of the energy spectrum given in Table 1 for: (a) $\ell = 0$, and (b) $\ell = 1$. The horizontal axis is the radial coordinate measured in units of the Bohr radius.

**Fig. 2**: The un-normalized wavefunctions in the finite basis (6) as given analytically by the series (14) shown in blue super-imposed by those calculated numerically in the complete basis (16) in red. We took $p = \eta q = 5$, $Q = 2$, and $\ell = 1$.

**Fig. 3**: The un-normalized wavefunctions in the finite basis (6) as given analytically by the series (14) shown in blue super-imposed by those calculated numerically in the complete basis (16) in red. We used the parameter values of Table 2 with state plots corresponding to $m = 1$ and $\gamma = 1.973702871$.

**Table 1**

| k | $\ell = 0$ | $\ell = 1$ | $\ell = 2$ | $\ell = 3$ |
|---|---|---|---|---|
| 0 | 1.689836518 | 0.251015112 | 0.050288629 | 0.011582200 |
| 1 | 0.337411507 | 0.084819527 | 0.021878946 | 0.005967795 |
| 2 | 0.123110800 | 0.038411932 | 0.011425659 | 0.003470897 |
| 3 | 0.058497441 | 0.020566781 | 0.006704777 | 0.002194508 |
| 4 | 0.032357265 | 0.012276277 | 0.004266330 | 0.001474931 |
| 5 | 0.019771611 | 0.007908751 | 0.002881243 | 0.001038673 |
| 6 | 0.012964075 | 0.005391547 | 0.002036676 | 0.000758897 |
| 7 | 0.008960226 | 0.003839389 | 0.001492591 | 0.000571260 |

**Table 2**

| $m$ | $\gamma$ | $-E_0$ | $-E_1$ | $-E_2$ | $-E_3$ |
|---|---|---|---|---|---|
| 0 | 0.904753862 | 0.088604373 | 0.044160252 | 0.026297735 | 0.017415888 |
|   | 2.351197519 | 0.041574253 | 0.025056330 | 0.016731035 | 0.011957498 |
|   | 3.172671646 | 0.027927569 | 0.018272110 | 0.012878292 | 0.009562772 |
|   | 4.073926412 | 0.019185950 | 0.013415538 | 0.009904961 | 0.007611586 |
| 1 | 1.973702871 | 0.050720257 | 0.029200677 | 0.018946465 | 0.013277165 |
|   | 3.095983541 | 0.028917932 | 0.018793782 | 0.013185974 | 0.009759184 |
|   | 4.060300322 | 0.019286713 | 0.013474421 | 0.009942300 | 0.007636732 |
|   | 5.034941193 | 0.013648539 | 0.010052469 | 0.007710801 | 0.006101341 |
| 2 | 1.539965664 | 0.064078256 | 0.034819564 | 0.021813636 | 0.014931773 |
|   | 2.966289119 | 0.030706849 | 0.019723198 | 0.013729065 | 0.010103555 |
|   | 4.023558622 | 0.019562282 | 0.013635068 | 0.010043998 | 0.007705129 |
|   | 5.022510687 | 0.013704236 | 0.010087660 | 0.007734435 | 0.006117972 |

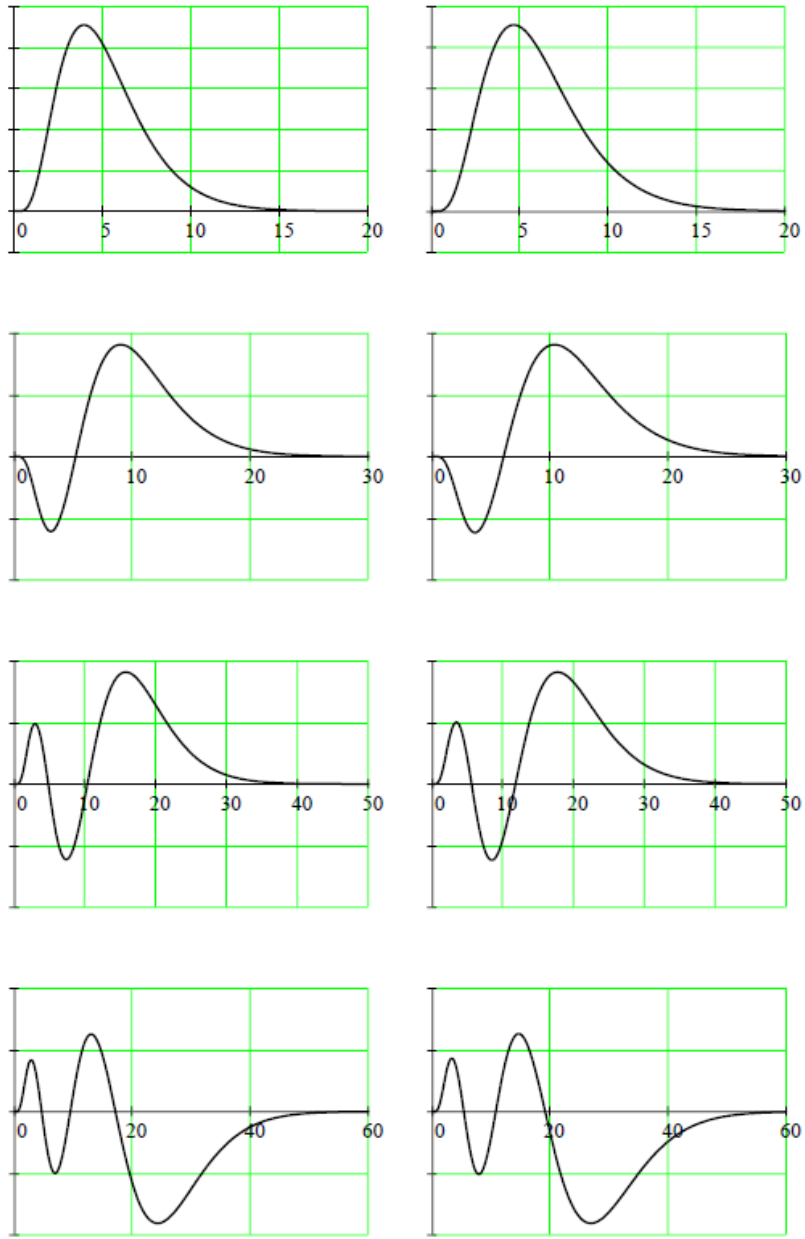

(a)          (b)

**Fig. 1**

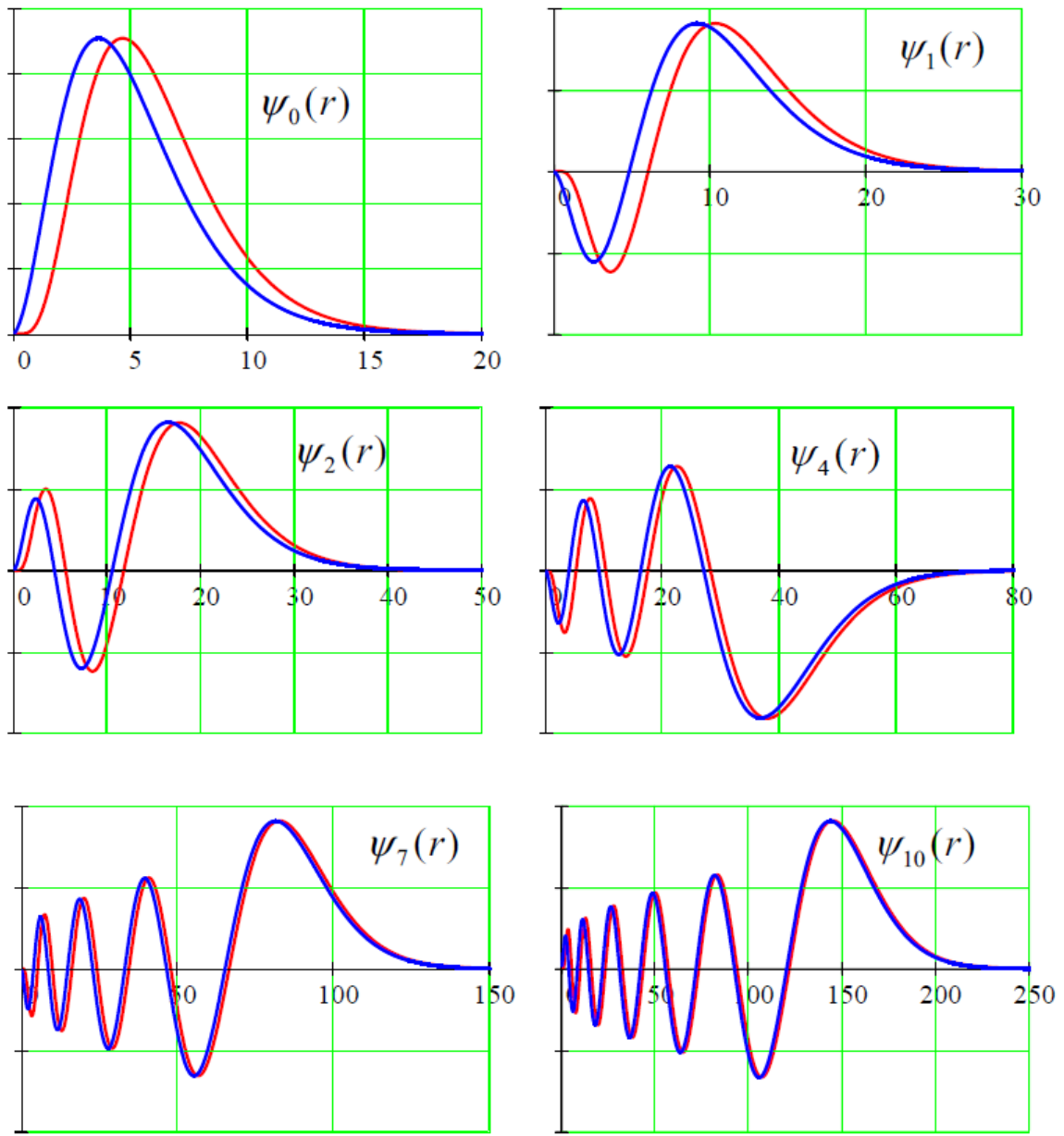

**Fig. 2**

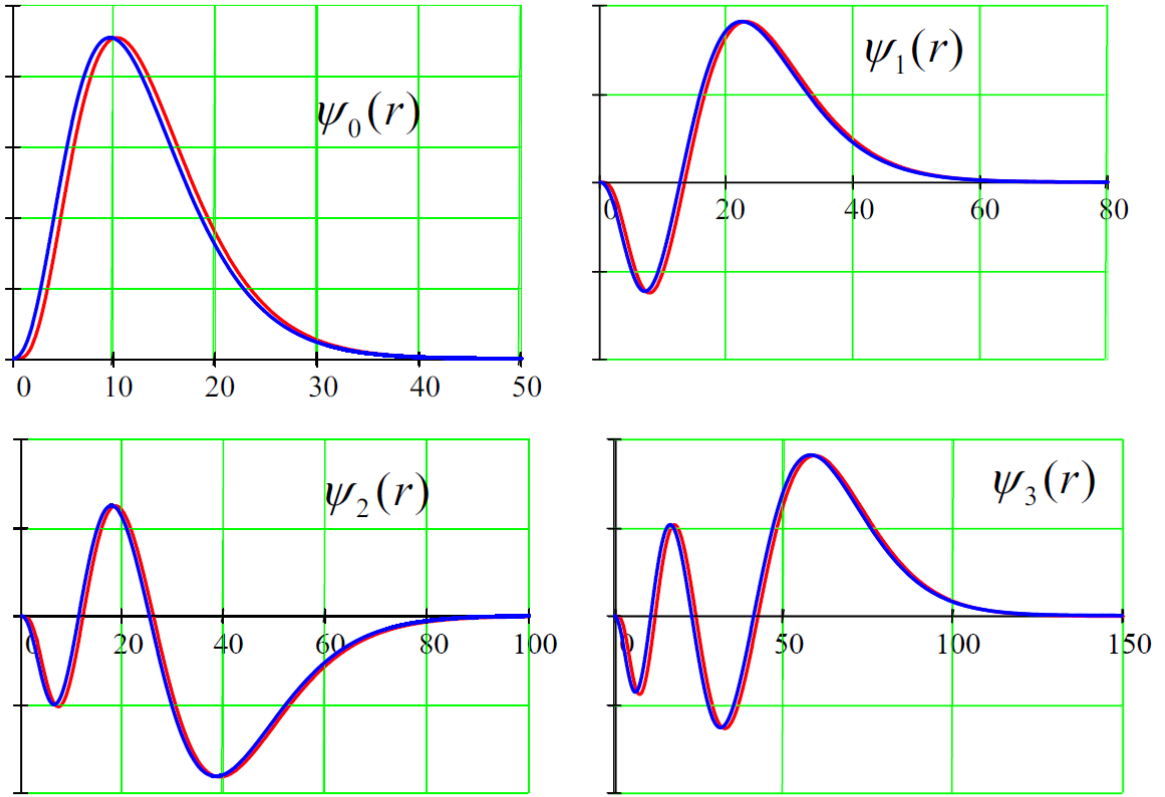

**Fig. 3**